\documentclass[12pt]{article}
\usepackage{wallpaper}
\usepackage[margin=1in, paperwidth=8.5in, paperheight=11in]{geometry}
\usepackage{amsmath}
\usepackage{graphicx}%
\usepackage{amsfonts}%
\usepackage{amssymb}
\usepackage{color}
\usepackage{float,subfigure}
\usepackage[round]{natbib}
\usepackage{cite}
\usepackage{setspace}
\usepackage{comment}
\usepackage{enumerate}
\usepackage{verbatim}
\usepackage{rotating}
\usepackage{microtype}
\usepackage[normalem]{ulem}
\usepackage{url}
\usepackage[hidelinks]{hyperref} 
\usepackage{cancel}
\usepackage{animate}
\usepackage{mathtools}
\usepackage{epsfig}
\usepackage{epstopdf}

\usepackage{amsthm}
\usepackage{multirow}

\hypersetup{pdfpagemode=UseNone}
\theoremstyle{plain}
\newtheorem{thm}{Theorem} 
\newtheorem*{thm*}{Theorem} 

\theoremstyle{definition}
\newtheorem{defn}[thm]{Definition} 

\newcommand{\appropto}{\mathrel{\vcenter{
  \offinterlineskip\halign{\hfil$##$\cr
    \propto\cr\noalign{\kern2pt}\sim\cr\noalign{\kern-2pt}}}}}

\newcommand{\bbeta}{ \ensuremath{\boldsymbol{\beta}}}

\newcommand{\sig}{ \ensuremath{\sigma}}

\newcommand{\bx}{ {\bf x} }

\newcommand{\by}{ {\bf y} }

\newcommand{\bz}{ {\bf z} }

\newcommand{\logit}{\mbox{logit}}

\newcommand{\given}{\,\vert\,}

\newcommand{\CAR}{\mbox{$\text{CAR}$}}

\newcommand{\tbeta}{\mbox{$\text{Beta}$}}

\newcommand{\IG}{\mbox{$\text{IG}$}}

\newcommand{\Bin}{\mbox{$\text{Bin}$}}

\newcommand{\N}{\mbox{Norm}}

\newcommand{\bpi}{ \mbox{\boldmath $ \pi $} }

\begin{document}

\thispagestyle{empty}
\setcounter{page}{0}
\singlespacing
\begin{center}
{\Large \textbf{Reliable event rates for disease mapping}} %

\bigskip

\textbf{Harrison Quick$^{*}$ and Guangzi Song}\\ 
Department of Epidemiology and Biostatistics, Drexel University, Philadelphia, PA 19104\\
$^{*}$ \emph{email:} hsq23@drexel.edu

\end{center}

\textsc{Summary.}
When analyzing spatially referenced event data, the criteria for declaring rates as ``reliable'' is still a matter of dispute. What these varying criteria have in common, however, is that they are rarely satisfied for crude estimates in small area analysis settings, prompting the use of spatial models to improve reliability.  While reasonable, recent work has quantified the extent to which popular models from the spatial statistics literature can overwhelm the information contained in the data, leading to oversmoothing.  Here, we begin by providing a definition for a ``reliable'' estimate for event rates that can be used for crude and model-based estimates and allows for discrete and continuous statements of reliability.  We then construct a spatial Bayesian framework that allows users to infuse prior information into their models to improve reliability while also guarding against oversmoothing. We apply our approach to county-level birth data from Pennsylvania, highlighting the effect of oversmoothing in spatial models and how our approach can allow users to better focus their attention to areas where sufficient data exists to drive inferential decisions.  We then conclude with a brief discussion of how this definition of reliability can be used in the design of small area studies.

\textsc{Key words:}
{Bayesian inference, Informative priors, Preterm birth, Small area analyses, Spatial statistics} 

\newpage

\doublespacing
\section{Introduction}
When producing estimates of the incidence of adverse health events (e.g., heart disease related deaths) or the prevalence of a risk factor (e.g., obesity) in small areas, there is a lack of consensus on what constitutes a ``reliable'' estimate. For instance, the United States Cancer Statistics (USCS) Working Group recommends declaring an estimate reliable if it is based on 16 or more cases \citeyearpar[USCS,][]{uscs}, while groups such as the New York Department of Health \citeyearpar{NY} require 20 or more cases.  As an illustration of the complexity of this issue, the Rhode Island Department of Health \citeyearpar{RI} and the Utah Department of Health \citeyearpar{UT} each provide detailed flowcharts describing when to report an estimate, when to report with a warning about reliability, and when to suppress estimates or aggregate data in their published reports.  More recently, the National Center for Health Statistics \citep[NCHS;][]{nchs:standards} produced a report detailing their new standards for reporting proportions, which consist of a mix of guidance based on the sample size (greater than 30) and the width of the estimate's confidence interval.  Many of these approaches are based on the \emph{coefficient of variation} (CV; also referred to as the \emph{relative standard error}) --- i.e., the standard error of the rate estimate divided by the estimate itself --- though the rationale underlying these rules is still quite vague.  An exception to this is the criteria of the USCS, which requires that the ratio of an estimate to the width of its 95\% confidence interval (assuming normality) be greater than 1, which is equivalent to requiring $\mbox{CV} < 1\slash 4$. Regardless of the criteria used, however, it is often the case that data from small areas will fail to satisfy the necessary criteria, especially when the data are stratified by demographic factors such as race/ethnicity, sex, and age.

One approach to relieve concerns about reporting estimates with high uncertainty is to aggregate the data across neighboring spatial regions and/or adjacent time periods.  For instance, the \emph{County Health Rankings and Roadmaps} program \citep[CHR\&R;][]{chr} aggregates data over periods of up to seven years in order to provide reliable estimates for as many counties as possible when constructing the measures used to determine their rankings.  While aggregation should boost case counts --- and thus lead to more reliable estimates --- it may also preclude inference on fine-level geographic disparities and/or temporal changes.  Perhaps worse, spatially aggregating neighboring regions with vastly different underlying rates can produce estimates that misrepresent each of the individual regions \citep[e.g.,][]{seth:aggregation,bradley:cage}. These concerns serve as a motivation for this work.

When analyzing spatially referenced data, an attractive alternative to data aggregation is to report model-based estimates generated in a Bayesian framework.  For instance, methods such as the conditional autoregressive (CAR) model of \citet{bym} and its multivariate extensions \citep[e.g.,][]{gelfand:mcar,quick:waller} have been used to produce over 40 years of county-level estimates of heart disease mortality by age, race, and sex \citep{adam:arg}, county-level estimates of suicide rates from 2005--2015 \citep{diba:suicide}, and census tract-level estimates of obesity rates \citep{quick:obese}.  Here, the benefit of using Bayesian methods is two-fold: not only do we obtain more precise (and thus more reliable) estimates via the infusion of prior information, but we may also obtain more \emph{accurate} estimates by virtue of Bayesian shrinkage \citep{stein1956}.  The key drawback of this, however, is that tasks such as quantifying and specifying the amount of information contributed by our models (relative to the data) is
not always straightforward \citep[e.g.,][]{morita:2008}.  For instance, recent work by \citet{bym:info} illustrated that the CAR model of \citeauthor{bym}\ could contribute the equivalent of more than 30 events per spatial region in the context of Poisson-distributed mortality data, and similar results were found by \citet{song:bin} in the context of binomially distributed birth data.  In both cases, the contribution of the CAR model framework --- when left to its own devices --- \emph{far} exceeded the contribution of the data in most regions, which can lead to oversmoothing and potentially untrustworthy inference.

The objectives of this paper are two-fold.  First and foremost, we provide a definition for a ``reliable'' estimate of an incidence or prevalence rate that is based on statistical point and interval estimates.  Not only does this definition accommodate crude and model-based estimates, but it also allows for discrete and continuous statements of reliability --- e.g., distinctions of \emph{reliable} versus \emph{unreliable} and the ability to indicate when one estimate is \emph{more reliable} than another.  Secondly, by anchoring this definition in a Bayesian framework, we allow users to infuse prior information (e.g., spatial structure) into their models to improve the reliability of their estimates.  In doing so, however, we must also be cognizant of the influence these models may have on our estimates.  As such, the foundation for the proposed work is the need to ``embrace uncertainty.''  More specifically, rather than designing models to ensure that \emph{all} estimates are reliable, we will provide guidance for restricting the informativeness of the commonly used \citeauthor{bym}~CAR model with an eye toward requiring a sufficient number of cases be observed for estimates to be deemed ``reliable.''

\section{Methods}\label{sec:methods}
\subsection{Definition of reliability}
We begin by assuming $y_i \sim \Bin\left(n_i,\pi_i\right)$, where $y_i$ denotes the number of cases in region $i$ out of a population (or in the case of survey data, a \emph{sample}) of size $n_i$ and where $\pi_i$ denotes the underlying event rate, for $i=1,\ldots,I$.  For the purpose of establishing our definition for a reliable event rate, we then assume $\pi_i \sim \tbeta\left(a_i,b_i\right)$, which yields a posterior of the form
\begin{align}
\pi_i \given y_i \sim \tbeta\left(y_i + a_i, n_i - y_i + b_i\right), \label{eq:post}
\end{align}
and yields interpretations of $a_i$ and $b_i$ as the prior number of cases and the prior number of non-cases, respectively, out of a total prior population of size $a_i+b_i$.
In this framework, the USCS's criteria for a reliable rate would require 
\begin{align}
\mbox{CV}\left(\pi_{i}\given y_{i}\right) = \frac{\sqrt{V\left[\pi_{i}\given y_{i}\right]}}{E\left[\pi_{i}\given y_{i}\right]} = \sqrt{\frac{n_i-y_i+b_i}{\left(y_i+a_i\right)\left(n_i+a_i+b_i+1\right)}} < 1\slash 4. \label{eq:cv}
\end{align}
To identify conditions for $y_i$ in which the posterior in~\eqref{eq:post} would yield a reliable estimate, we first note that the requirement in~\eqref{eq:cv} can be reexpressed as a requirement on the posterior number of cases:
\begin{align}
\left(y_i+a_i\right) > 16 \times \frac{n_i-y_i+b_i}{\left(n_i+a_i+b_i+1\right)} \approx 16 \left(1-E\left[\pi_{i}\given y_i\right]\right),\label{eq:rq}
\end{align}
when 
$n_i + a_i+b_i$ is large.  A convenient feature of the requirement in~\eqref{eq:cv} is that if we were to approximate the 95\% credible interval (CI) for $\pi_i$ as $E\left[\pi_{i}\given y_{i}\right] \pm 2 \sqrt{V\left[\pi_{i}\given y_{i}\right]}$, then satisfying~\eqref{eq:rq} would also result in $E\left[\pi_{i}\given y_{i}\right]$ being greater than the width of its 95\% CI.  Finally, while the relationships in~\eqref{eq:post}--\eqref{eq:rq} are based on \emph{binomially} distributed data, similar results hold when the data are modeled as being \emph{Poisson} distributed, as is customary when analyzing birth and death rates \citep{brillinger} and other rare event data.  Analogous derivations under the Poisson model specification are provided in Web Appendix~A, the requirements under which are consistent with those for binomial data when $\pi$ is small.

We thus generalize the USCS reliability criteria as a function of quantiles of the posterior distribution as follows:
\begin{defn}\label{dfn:reliable}
\emph{Estimates of the rate parameter $\pi$ obtained from the posterior distribution $p\left(\pi\given \by\right)$ are \emph{reliable at the $1-\alpha$ level} if the posterior medians of $\pi$ and its opposite, $1-\pi$, are each larger than the width of their respective $(1-\alpha)\times 100$\% equal-tailed credible intervals.}
\end{defn}
\noindent 
We also define the \emph{relative precision} of an estimate at the $1-\alpha$ level to be the ratio of its posterior median over the width of its $(1-\alpha)\times 100$\% credible interval, where a relative precision greater than 1 corresponds to an estimate that is reliable at the $1-\alpha$ level.  Finally, note that Definition~\ref{dfn:reliable} is designed to ensure that $\pi$ and its opposite, $1-\pi$, receive the same reliability distinction --- e.g.,
the estimate for the prevalence of people who \emph{do} smoke has the same level of reliability as the estimate for the prevalence of people who \emph{do not} smoke.

\subsection{Impact of informative priors}\label{sec:impact}
\begin{figure}[t]
    \begin{center}
        \subfigure[Relative Precision $\times$ Event Rates]{\includegraphics[width=.45\textwidth]{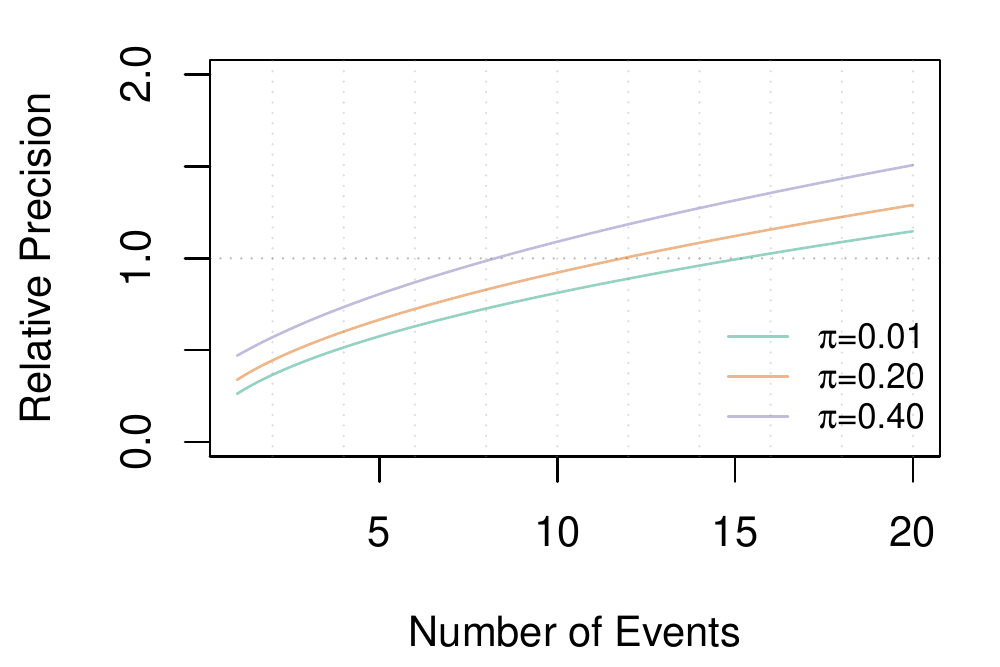}\label{fig:relcurvea}}
        \subfigure[Relative Precision $\times$ Levels of Reliability]{\includegraphics[width=.45\textwidth]{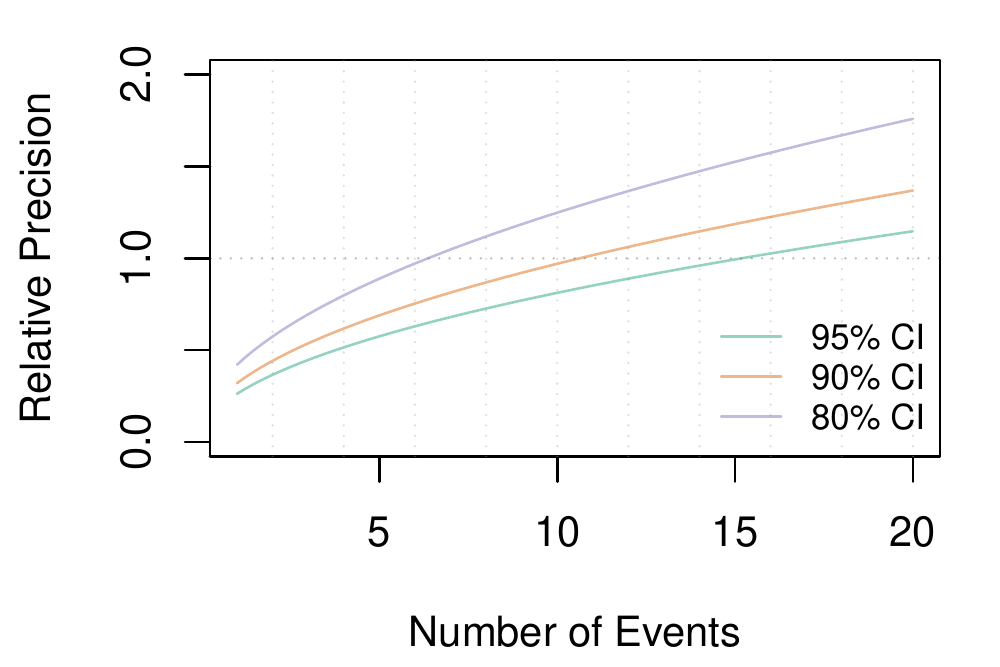}\label{fig:relcurveb}}
    \end{center}
    \caption{Comparison of the relative precision as a function of the number of events.  Panel~(a) displays the relative precision at the 0.95 level for various underlying event rates, and Panel~(b) displays the relative precision for various levels of reliability for an event rate of $\pi_i=0.01$.}
    \label{fig:curves}
\end{figure}

Using this definition of reliability, the requirement from~\eqref{eq:rq}, and a relatively noninformative prior for $\pi_i$ --- i.e., $a_i=1\slash 2$ and $b_i=a_i \left(1-\pi_{i0}\right)\slash \pi_{i0}$ such that $\pi_{i0}=a_i\slash\left(a_i+b_i\right)$ --- we would need to observe 16 cases to obtain a reliable estimate at the 0.95 level when $\pi_i=0.01$, 12 cases when $\pi_i=0.20$, and 9 cases when $\pi_i=0.40$, as shown in Figure~\ref{fig:relcurvea}.  Similarly, if we were to relax the desired level of reliability and assume $\pi_i=0.01$, approximately 16 events would be required to deem a rate reliable at the 0.95 level, 11 events at the 0.90 level, and 7 events at the 0.80 level, as displayed in Figure~\ref{fig:relcurveb}.  From this point forward, references to reliability and the relative precision are at the 0.95 level unless otherwise stated.

While the results above are convenient, the true benefit of this definition of reliability is revealed when we consider \emph{informative} prior specifications for $\pi_i$. For instance, suppose we are analyzing a dataset comprised of the number of infants born preterm (i.e., before 37 weeks of pregnancy) 
out of the total number of births in one or more small areas and that prior information indicates that 10\% of infants are born preterm.  Based on~\eqref{eq:rq}, our posterior number of preterm births, $y_i+a_i$, must exceed $16\times\left(1-0.10\right)=14.4$ for the estimates from these areas to be deemed ``reliable''.  Thus, rather than \emph{explicitly} requiring $y_i$ to exceed a given threshold for an estimate to be deemed reliable, we can instead view such requirements in terms of a maximum on the amount of information that can be contributed by the prior.  For instance, if $a_i < 16 \times \left(1-0.10\right)\slash 2 = 7.2$, then the data must contribute more information to our estimate than the prior --- i.e., $y_i > a_i$ --- in order for an estimate to be deemed reliable. 

That said, \emph{explicitly} incorporating prior information as described above is not a common practice.  Furthermore, in the context of disease mapping, it is common to consider a model specification in which
\begin{align}
\logit\left(\pi_i\right) \given \bbeta,\bz,\sig^2 \sim \N\left(\bx_i^T\bbeta + z_i,\sig^2\right), \label{eq:logit}
\end{align}
where $\bx_i$ denotes a $p$-vector of region-specific covariates with a corresponding vector of regression coefficients, $\bbeta$, and $z_i$ is a spatial random effect based on the conditional autoregressive (CAR) model framework of \citet{bym}, denoted by $\bz\sim\CAR\left(\tau^2\right)$.  While measuring the informativeness of the model specification in~\eqref{eq:logit} is nontrivial, recent work by \citet{song:bin} established a relationship between between the posterior in~\eqref{eq:post} and the posterior resulting from~\eqref{eq:logit}.  This was achieved by first equating the mean and the variance of a beta distribution and a logitnormal using the delta method, and then extending that relationship to the conditional distribution of $\pi_i$ given the remaining $\pi_j$, $j\ne i$, under~\eqref{eq:logit} after integrating the spatial random effects, $\bz$, out of the model, which yielded an approximation of the form: 
\begin{align}
\widehat{a}_i =\frac{1+\exp(\bx_i^T\bbeta)}{\sigma^2+\left(\sigma^2+\tau^2\right)/m_i}-\frac{\exp(\bx_i^T\bbeta)}{1+\exp(\bx_i^T\bbeta)}, \label{eq:info}
\end{align}
where $m_i$ denotes the number of regions that neighbor region $i$.  Because each region can have its own covariate vector, $\bx_i$, and its own number of neighbors, $m_i$, \citet{song:bin} recommended defining the model's baseline level of informativeness as
\begin{align}
\widehat{a}_0 =\frac{1+\exp(\bx_0^T\bbeta)}{\sigma^2+\left(\sigma^2+\tau^2\right)/m_0}-\frac{\exp(\bx_0^T\bbeta)}{1+\exp(\bx_0^T\bbeta)}, \label{eq:bin_info}
\end{align}
wherein $\bx_i$ and $m_i$ from~\eqref{eq:info} are replaced with $\bx_0$ and $m_0$, respectively, where $\bx_0=\left(x_{01},\ldots,x_{0p}\right)'$ represents the vector of the average covariate values --- i.e., $x_{0j} = \sum_i x_{ij}\slash I$ for $j=1,\ldots,p$ --- and $m_0$ represents a baseline number of neighbors with $m_0=3$ suggested as a rule-of-thumb to allow for comparisons between datasets.  Following a similar process, \citet{bym:info} estimated the informativeness of the analogous Poisson model specification as
\begin{align}
\widehat{a}_0 = \frac{1}{\exp\left[\sigma^2+\left(\sigma^2+\tau^2\right)/m_0\right] -1}.\label{eq:pois_info}
\end{align}
A brief description of how the model in~\eqref{eq:logit} can be fit under a restriction that $\widehat{a}_0 < A$ for some $A>0$ is provided in Web Appendix~B.

\section{Case Study: Rates of Preterm Birth in PA Counties}\label{sec:analysis}
To illustrate how this approach can be used in practice, we conduct a case study pertaining to preterm birth in Pennsylvania counties from 2010--2019.  The data are stratified by both year and race/ethnicity of the mother (white, black, Hispanic, and Asian) and were obtained from the Pennsylvania Department of Health's web-based Enterprise Data Dissemination Informatics Exchange (EDDIE) system \citep{eddie}.  In the EDDIE system, race/ethnicity is coded as white, black, Hispanic, or Asian with the caveat that these groups may not be mutually exclusive (e.g., mothers coded as ``white'' may be Hispanic or non-Hispanic white).  For the purposes of our analysis, however, we will analyze each of these groups separately.  Similarly, while it may be attractive to consider the use of models that account for correlation between mothers of different racial/ethnic backgrounds and/or temporal relationships in the underlying incidence rates (e.g., by using the multivariate spatiotemporal CAR model structure of \citet{nonsep:poisson}), analyzing these scenarios separately will serve as a means of demonstrating replicability.

\begin{table}[t]
\begin{center}
\begin{tabular}{|l|c|c|c|c|}
\hline
 & White & Black & Asian & Hispanic\\
\hline
Average Births per Year & 97,934 & 19,848 &  5,970 & 14,809\\
Average Preterm Births per Year & 8,550 & 2,618 &  470 & 1,474\\
Average Preterm Birth Rate & 8.7\% & 13.2\% &  7.9\% & 10.0\% \\
Percent of Counties with $<$ 10 Preterm Births & 4.5\% & 67.2\% & 83.7\% & 70.4\% \\
Percent of Counties with Zero Preterm Births & 0.3\% & 38.4\% & 50.3\% & 31.3\% \\
Ratio of Urban vs.\ Rural Birth Totals & 0.44 & 3.27 & 1.95 & 0.64\\
\hline
\end{tabular}
\end{center}
\caption{Summary of the Pennsylvania preterm birth data, presented as averages over the ten-year period, 2010-2019.  Ratio of urban vs.\ rural births calculated by declaring counties with population densities greater than 1,000 people per square mile as ``urban'', which corresponds to the metropolitan areas of Philadelphia (Philadelphia County and the neighboring Bucks, Delaware, and Montgomery Counties) and Pittsburgh (Allegheny County).}
\label{tab:ptb}
\end{table}

A summary of the data is provided in Table~\ref{tab:ptb}.  Here, we see evidence of several important phenomena.  First and foremost, we see evidence of Pennsylvania's racial demographics, where not only were approximately 70\% of the state's births to white mothers, but also that racial/ethnic minority mothers --- particularly black and Asian mothers --- are more geographically concentrated in the state's urban centers, Philadelphia and Pittsburgh.  As a result, data from racial/ethnic minority mothers are sparse, with more than 30\% of Pennsylvania's counties experiencing \emph{zero} preterm births from black, Asian, and Hispanic mothers and two-thirds of counties experiencing fewer than 10.  That being said, Table~\ref{tab:ptb} also displays evidence of high racial/ethnic disparities in the incidence of preterm birth, with black mothers in Pennsylvania experiencing rates nearly 50\% higher than their non-black counterparts.  Thus, demographic challenges notwithstanding, exploring county-level trends in preterm birth \emph{by race} is of epidemiologic interest, thus motivating the use spatial models to produce more stable estimates than the data alone can provide. 

With this background in mind, we let $y_{irt}$ and $n_{irt}$ denote the number of preterm births and total number of births, respectively, to mothers of race $r$ in county $i$ in year $t$.  We then assume $y_{irt}\given \pi_{irt} \sim \Bin\left(n_{irt},\pi_{irt}\right)$ and implement two model specifications analogous to~\eqref{eq:logit} to model $\pi_{irt}$, stratified by each combination of race and year.
First, we will let
\begin{align}
p\left(\bpi_{\cdot rt},\beta_{0;rt},\bz_{\cdot rt},\sig_{rt}^2,\tau_{rt}^2\given \by_{\cdot rt}\right)\! \propto& \prod_{i=1}^{I}\left[\Bin\left(y_{irt}\given n_{irt},\pi_{irt}\right) \times \N\left(\logit\; \pi_{irt}\given \beta_{0;rt}+z_{irt},\sig_{rt}^2\right)\right] \notag \times\\
&\!\times \CAR\!\left(\bz_{\cdot rt}\given \tau_{rt}^2\right) \times \IG\!\left(\sig_{rt}^2\given 1,1\slash 100\right) \times \IG\!\left(\tau_{rt}^2\given 1,1\slash 7\right), \label{eq:standard}
\end{align}
denote a \emph{standard} CAR modeling approach, where the priors above for $\sig_{rt}^2$ and $\tau_{rt}^2$ are consistent with previous studies \citep[e.g.,][]{bernardinelli,waller:carlin} and the prior for $\beta_{0;rt}$ (i.e., a flat, improper prior) can be expressed as $p\left(\beta_{0;rt}\right) \propto 1$.  As the approach in~\eqref{eq:standard} may have a tendency to produce overly informative models, we also consider a model specification in which we restrict $\widehat{a}_{0;rs} < 5$; as discussed in Section~\ref{sec:impact}, this restriction should be expected to result in a requirement that $y_{irt}\ge 10$ for an estimate of $\pi_{irt}$ to be deemed reliable at the 0.95 level.
To achieve this, we follow the approach of \citet{song:bin} and let
\begin{align}
p\left(\bpi_{\cdot rt},\beta_{0;rt},\bz_{\cdot rt},\sig_{rt}^2,\tau_{rt}^2\given \by_{\cdot rt}\right) \propto& \prod_{i=1}^{I}\left[\Bin\left(y_{irt}\given n_{irt},\pi_{irt}\right) \times \N\left(\logit \pi_{irt}\given \beta_{0;rt}+z_{irt},\sig_{rt}^2\right)\right] \notag \times\\
&\times \CAR\left(\bz_{\cdot rt}\given \tau_{rt}^2\right) \times \IG\left(\sig_{rt}^2\given 1,1\slash 100\right) \notag\\
&\times \IG\left(\tau_{rt}^2\given 1,1\slash 7\right) \times I\left\{\widehat{a}_{0;rt} < 5\right\}, \label{eq:restrict}
\end{align}
denote our \emph{restricted} CAR modeling approach. 
Both the standard and restricted models were fit in {\tt R} \citep{R} using Markov chain Monte Carlo (MCMC) algorithms run for a total of 100,000 iterations, with separate runs for each combination of race/ethnicity and year.  For the sake of uniformity and to ease the computational burden of the \emph{post hoc} analyses, the first 50,000 iterations of each chain were discarded as burn-in and the last 50,000 iterations were thinned by a factor of 10, resulting in 5,000 iterations' worth of samples for each combination of race/ethnicity and year for both the standard and restricted models.

\begin{figure}[t]
    \begin{center}
       \includegraphics[width=.9\textwidth]{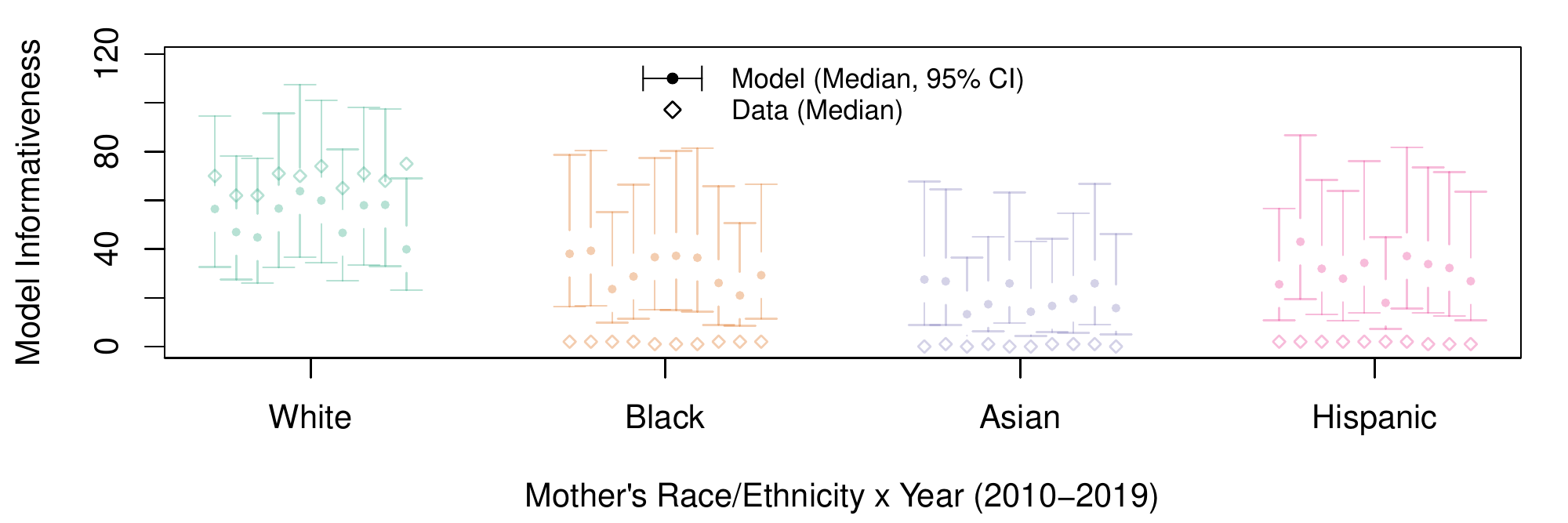}
    \end{center}
    \caption{Comparison of the estimated model informativeness parameters, $\widehat{a}_{0;rt}$, from the Pennsylvania preterm birth analysis; median county-level counts provided for reference.}
    \label{fig:info}
\end{figure}

Before we discuss the rate estimates themselves (and their degree of reliability), we consider the estimates for the informativeness of the standard CAR models (based on $\widehat{a}_{0;rt}$) shown in Figure~\ref{fig:info}.  Here, we see that the standard CAR model framework of \citet{bym} is contributing the equivalent of 13--43 preterm births per county per year for each of the racial minority groups, values that are far greater than the observed number of preterm births in most of those counties (as previously noted in Table~\ref{tab:ptb}).  While this is not quite the case for white mothers --- where the model appears to be contributing fewer preterm births per county than the data --- the model is still contributing in excess of 40 preterm births per county, thus estimates for all groups considered may be susceptible to oversmoothing.

\begin{figure}[t]
    \begin{center}
        \subfigure[White Mothers; McKean County]{\includegraphics[width=.45\textwidth]{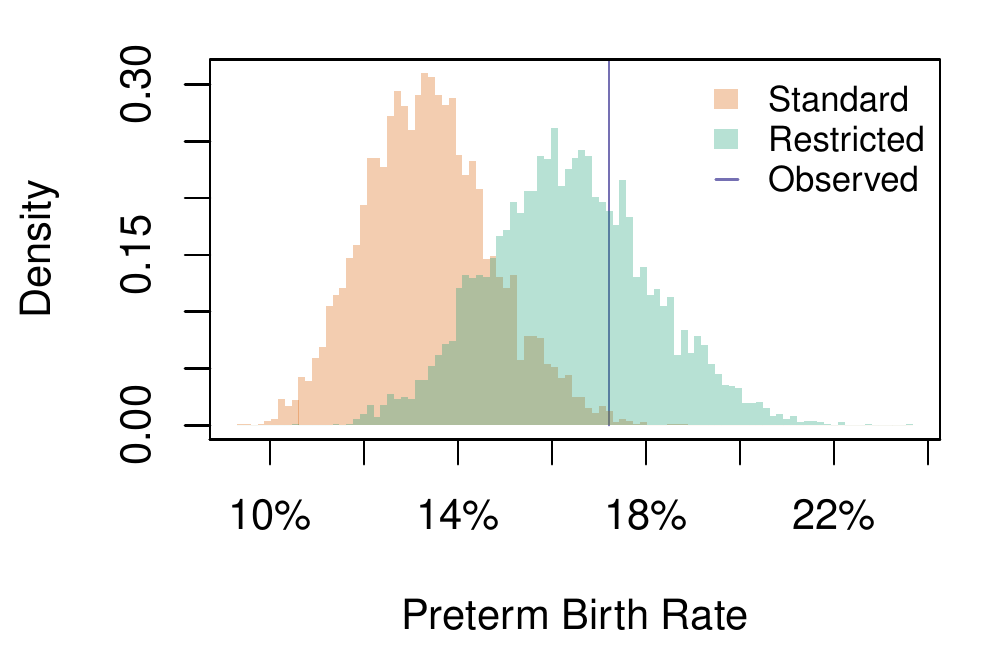}\label{fig:white_hist}}
        \subfigure[Black Mothers; Adams County]{\includegraphics[width=.45\textwidth]{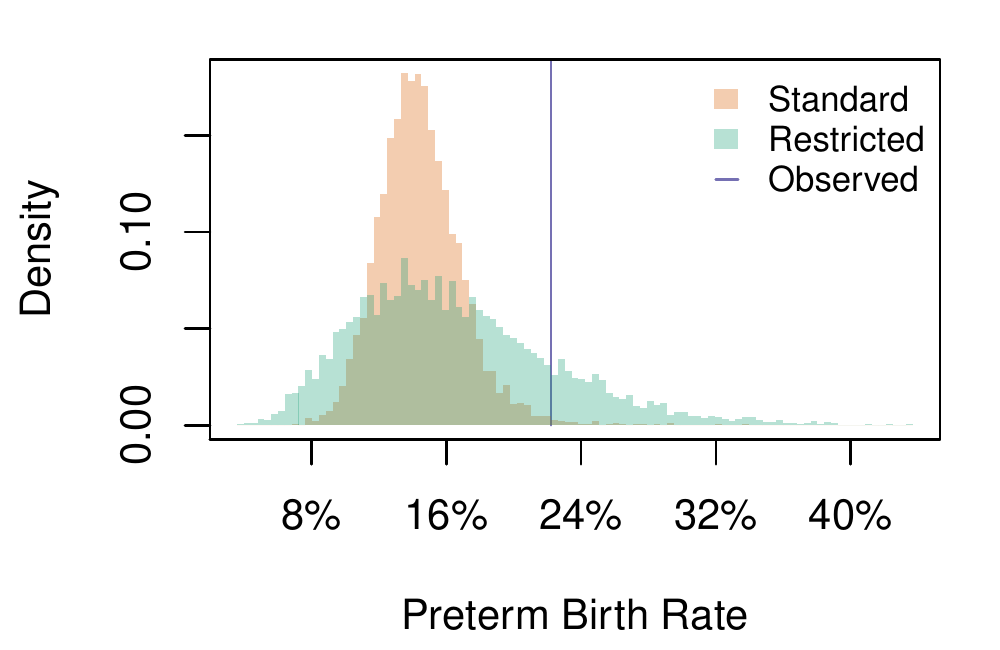}\label{fig:black_hist}}
    \end{center}
    \caption{Comparison of the posterior distributions of rate parameters, $\pi_{irt}$, under the standard and restricted CAR models for white and black mothers in selected counties.}
    \label{fig:histcomp}
\end{figure}

To illustrate the potential impacts of oversmoothing, we consider the estimates for white mothers in McKean County in 2010 and the estimates for black mothers in Adams County in 2019 shown in Figure~\ref{fig:histcomp}.  In Figure~\ref{fig:white_hist}, we see that while the restricted CAR model yields an estimate for white mothers in McKean County that is fairly consistent with the observed data (17.2\%) --- reflecting the relatively large number of preterm births observed in the data (74) --- the heightened informativeness of the standard CAR model has pulled the estimate away from its observed rate toward the rate in the neighboring counties (11.4\%).  As we illustrate at greater length in Web Appendix~C (and specifically in Figure~C.1), such extreme oversmoothing could inhibit our ability to detect outlying regions and thus could stymie state and local health departments' intervention efforts. Meanwhile, Figure~\ref{fig:black_hist} illustrates a second drawback of the standard CAR model's tendency to oversmooth estimates: an (unwarranted) increase in precision.  Specifically, Adams County is a predominantly white, rural county in which only two preterm births were observed for black mothers (out of nine total births) in 2019.  While the observed rate for this county is consistent with both of its neighbors and the overall state-level average for black mothers, the standard CAR model produces an estimate whose relative precision is twice that of the estimate produced by the restricted CAR model and consistent with that of a county with over 30 preterm births.

\begin{figure}[t]
    \begin{center}
        \subfigure[White Mothers: 2019 (Standard)]{\includegraphics[width=.4\textwidth]{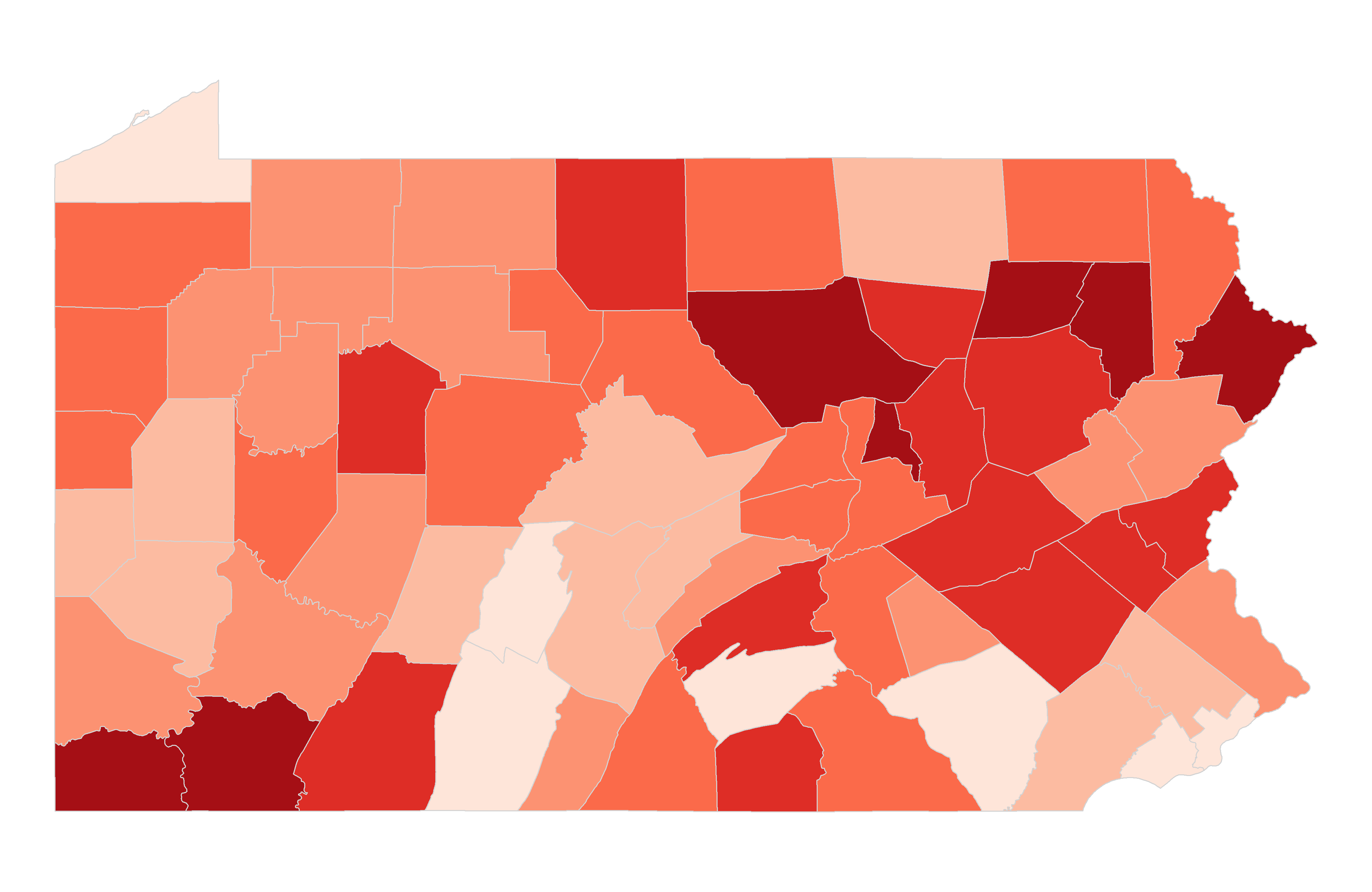}\label{fig:white_rate_full}}
        \subfigure[White Mothers: 2019 (Restricted)]{\includegraphics[width=.4\textwidth]{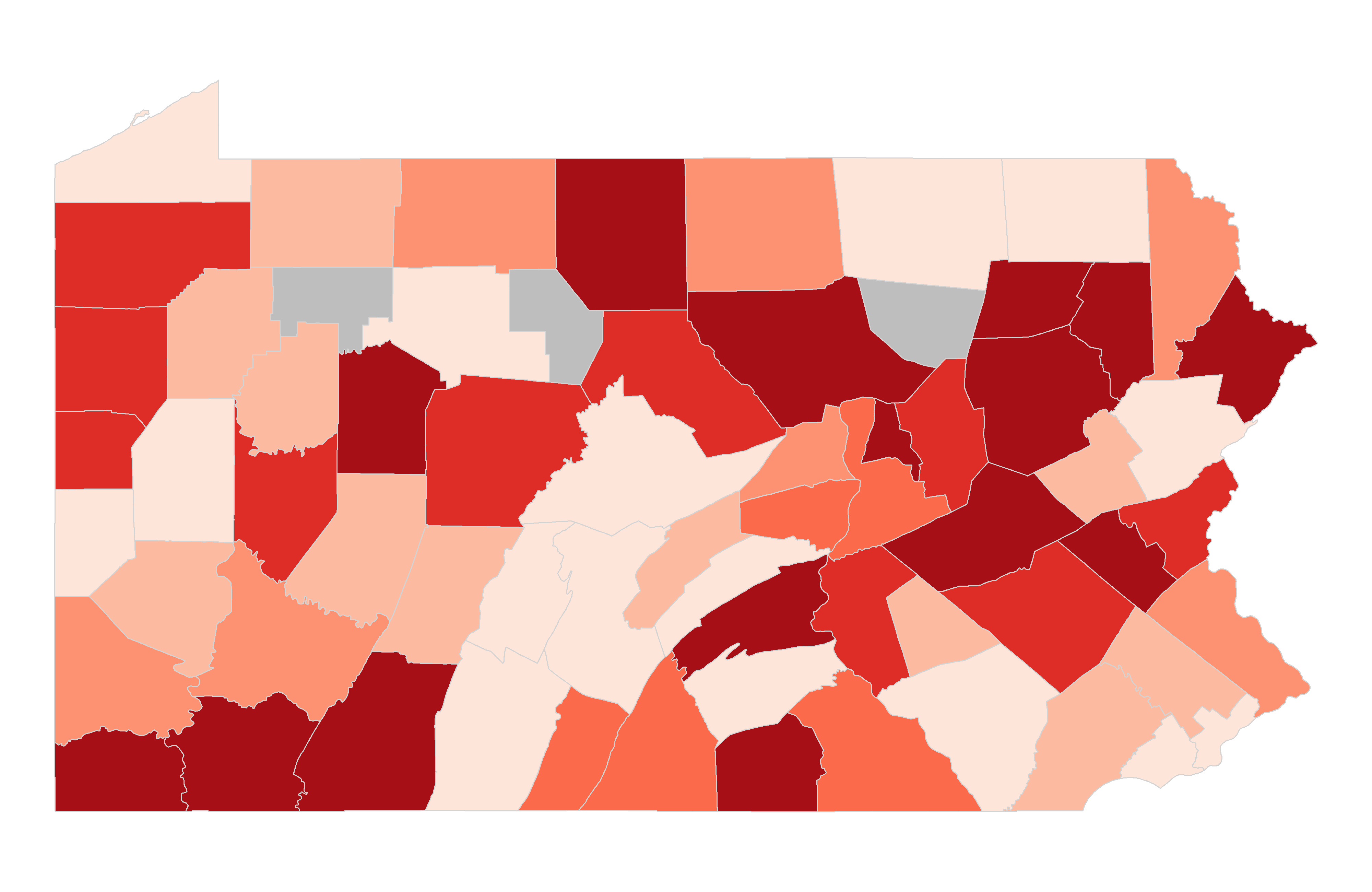}\label{fig:white_rate_thres}}
        \includegraphics[width=.16\textwidth]{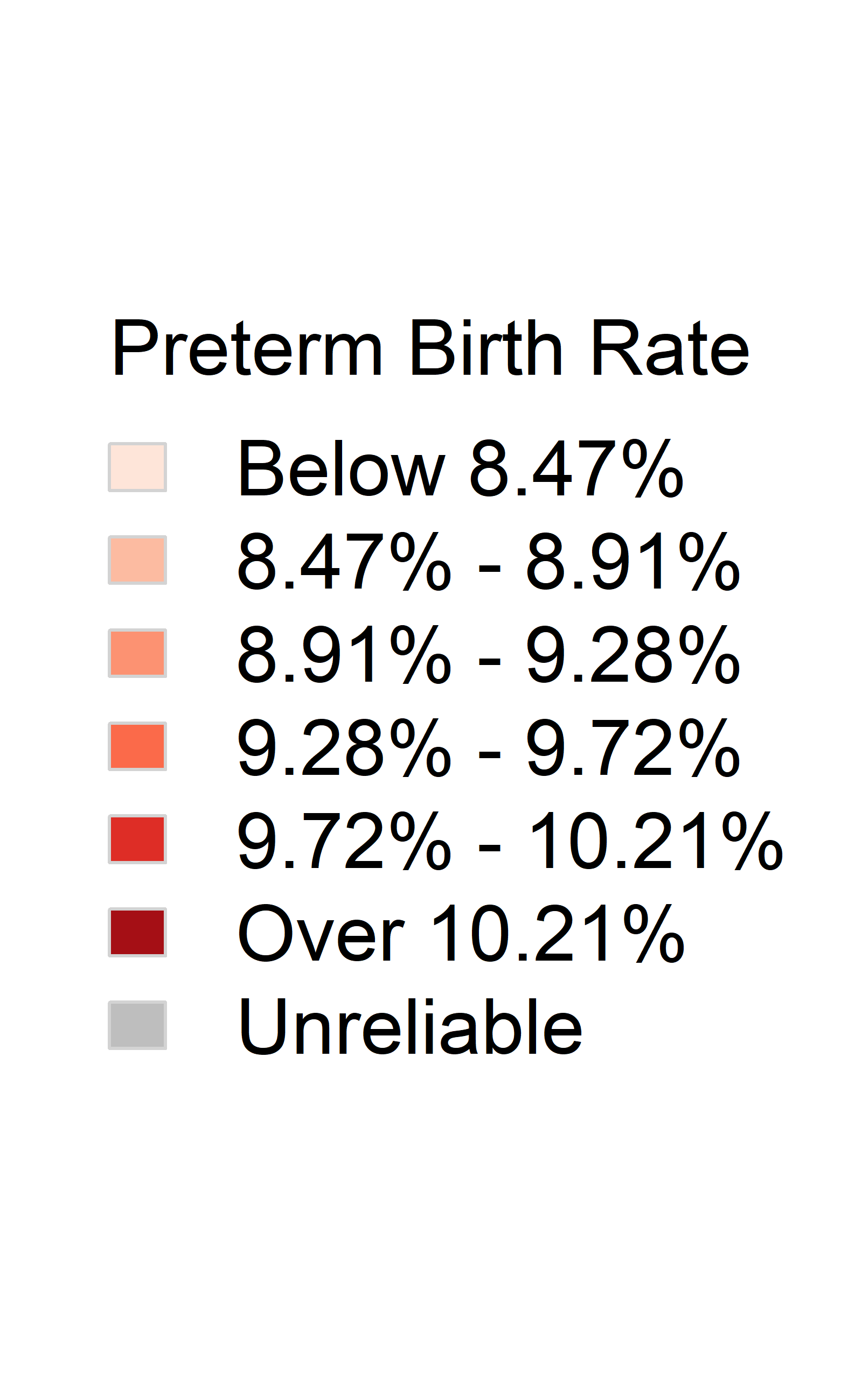}\\
        \subfigure[Asian Mothers: 2019 (Standard)]{\includegraphics[width=.4\textwidth]{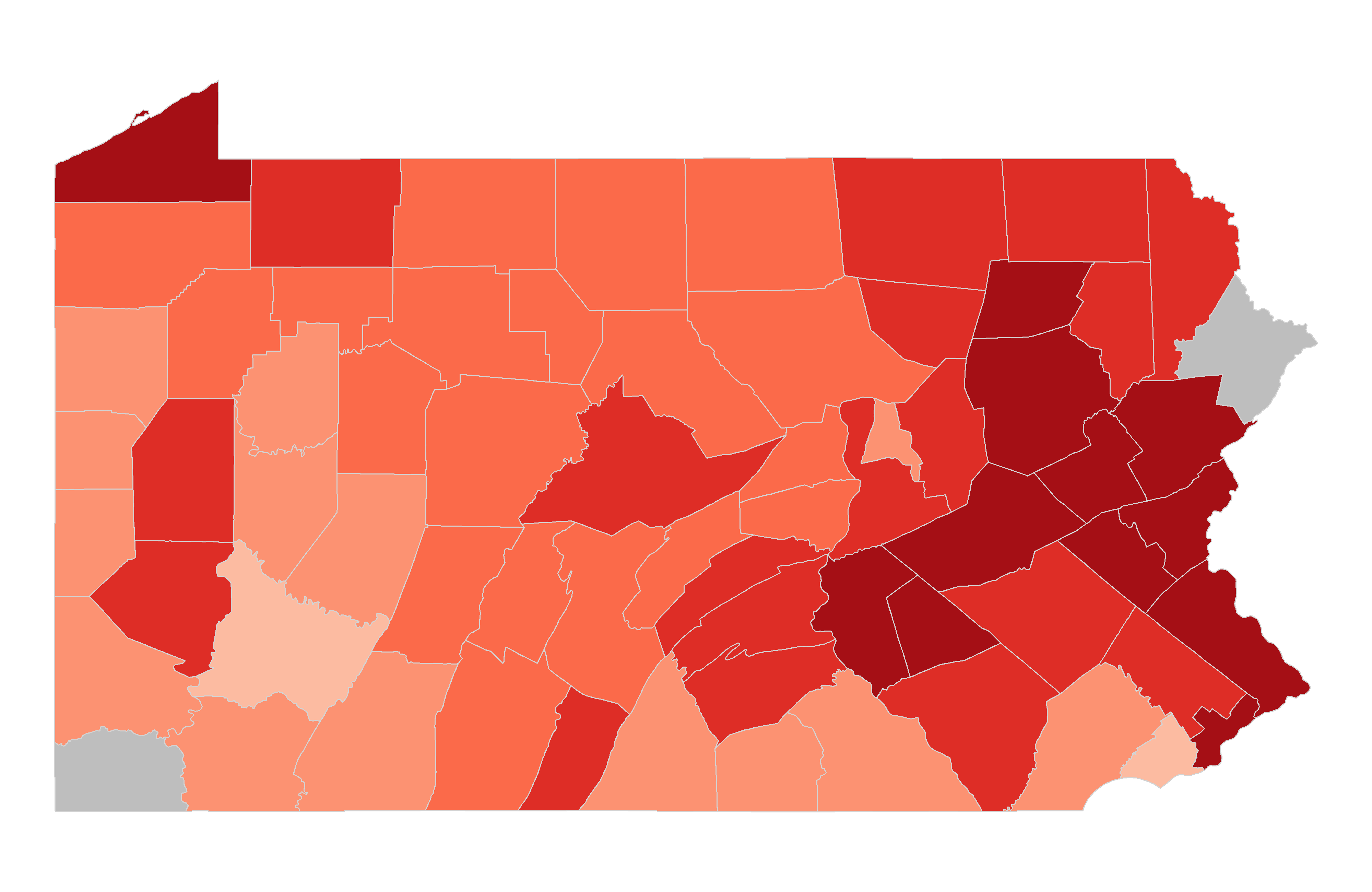}\label{fig:asian_rate_full}}
        \subfigure[Asian Mothers: 2019 (Restricted)]{\includegraphics[width=.4\textwidth]{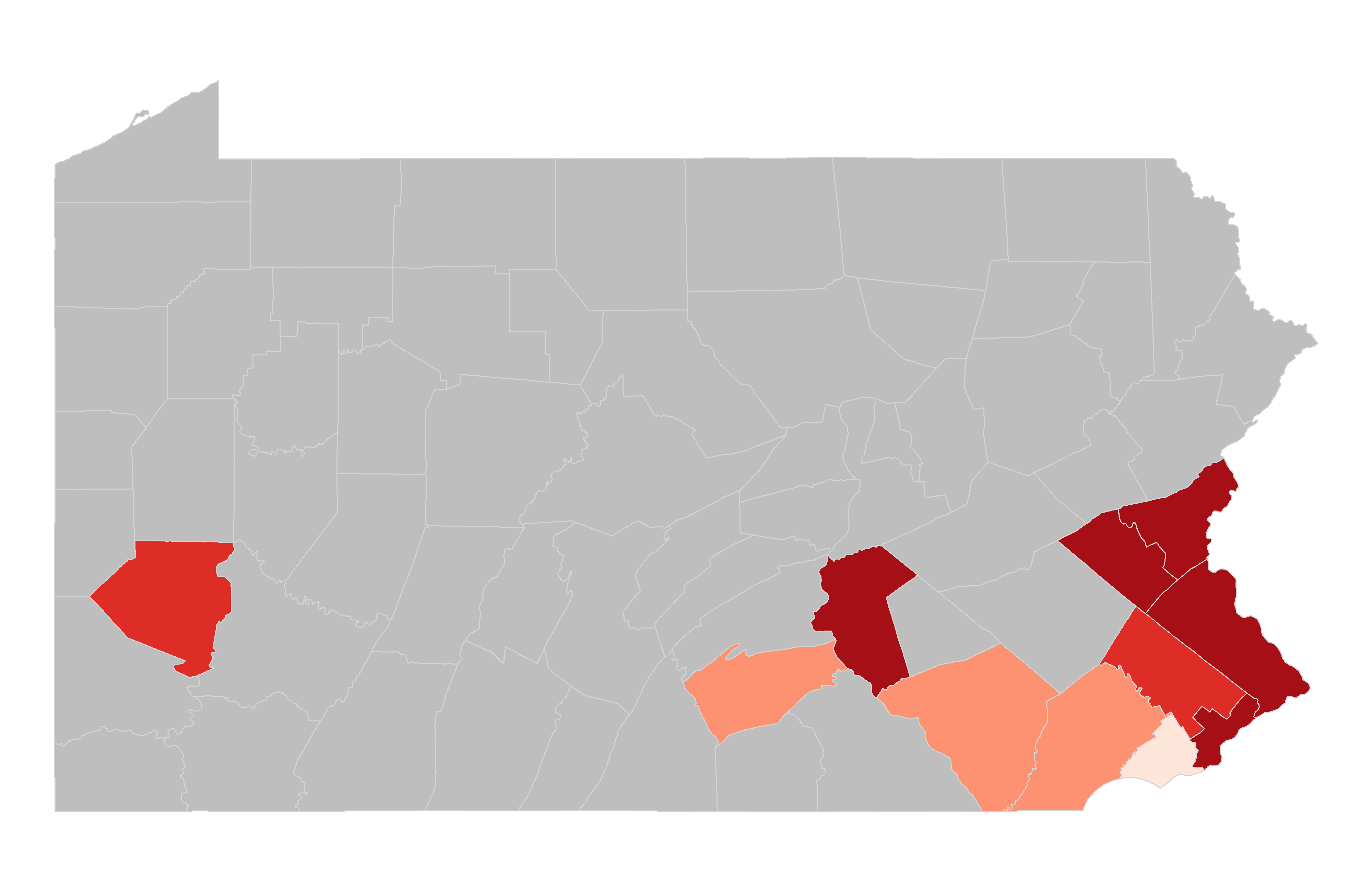}\label{fig:asian_rate_thres}}
        \includegraphics[width=.16\textwidth]{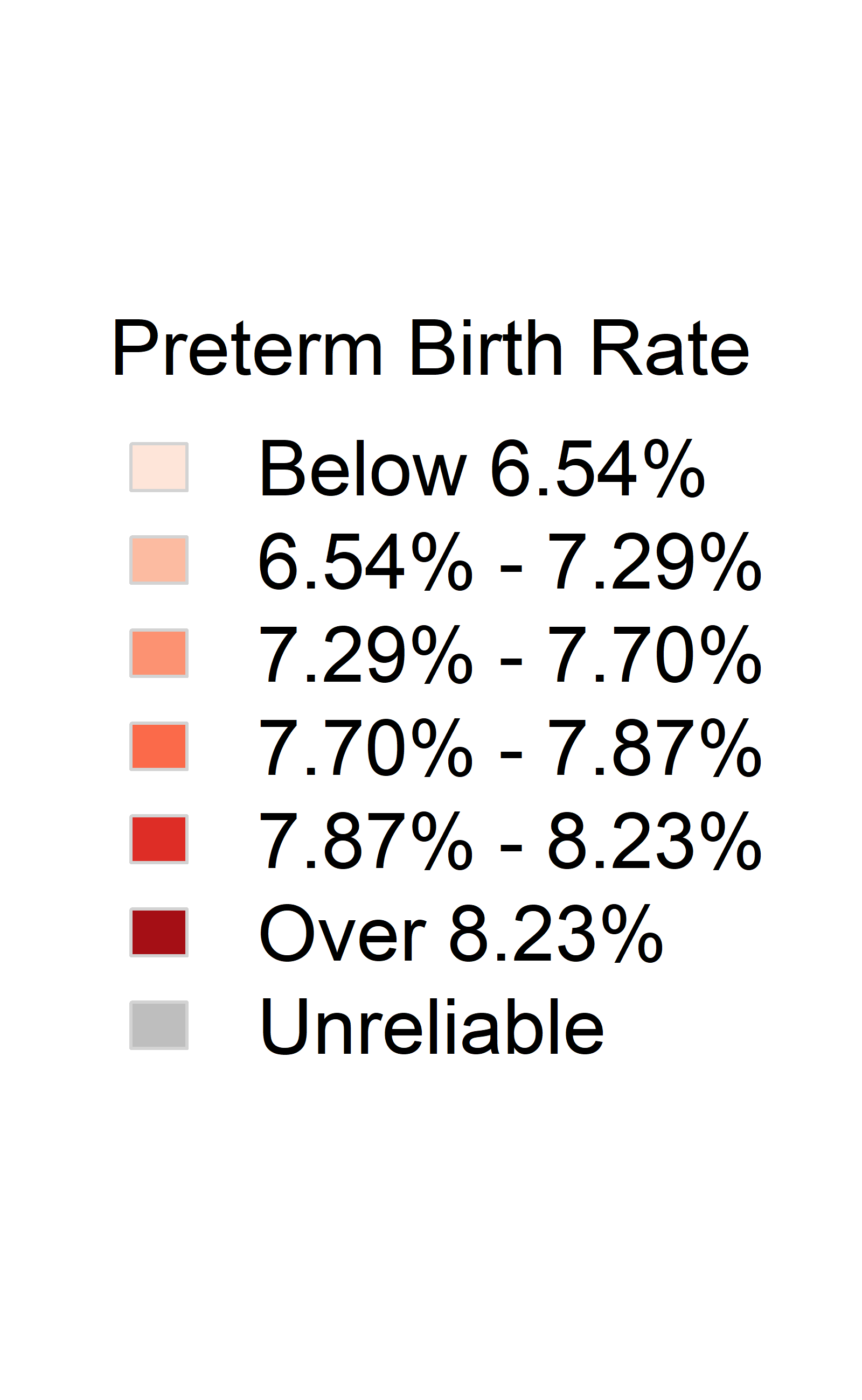}
    \end{center}
    \caption{Comparison of the preterm birth rates for white and Asian mothers in 2019 from the standard and restricted CAR models.}
    \label{fig:ratemaps}
\end{figure}

\begin{figure}[t]
    \begin{center}
        \subfigure[White Mothers: 2019]{\includegraphics[width=.45\textwidth]{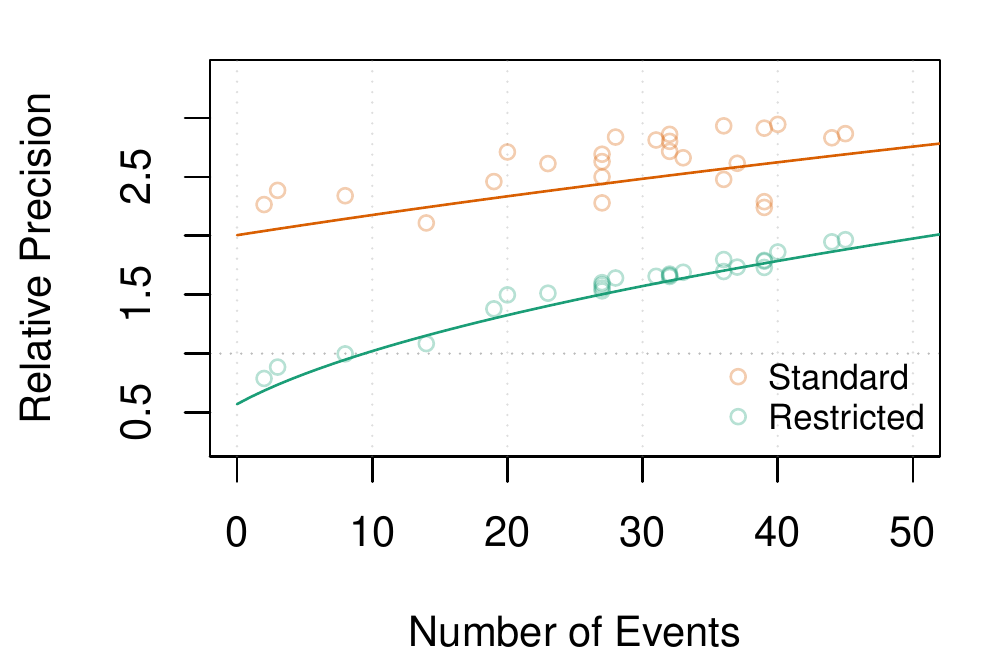}\label{fig:rprec_white}}
        \subfigure[Black Mothers: 2019]{\includegraphics[width=.45\textwidth]{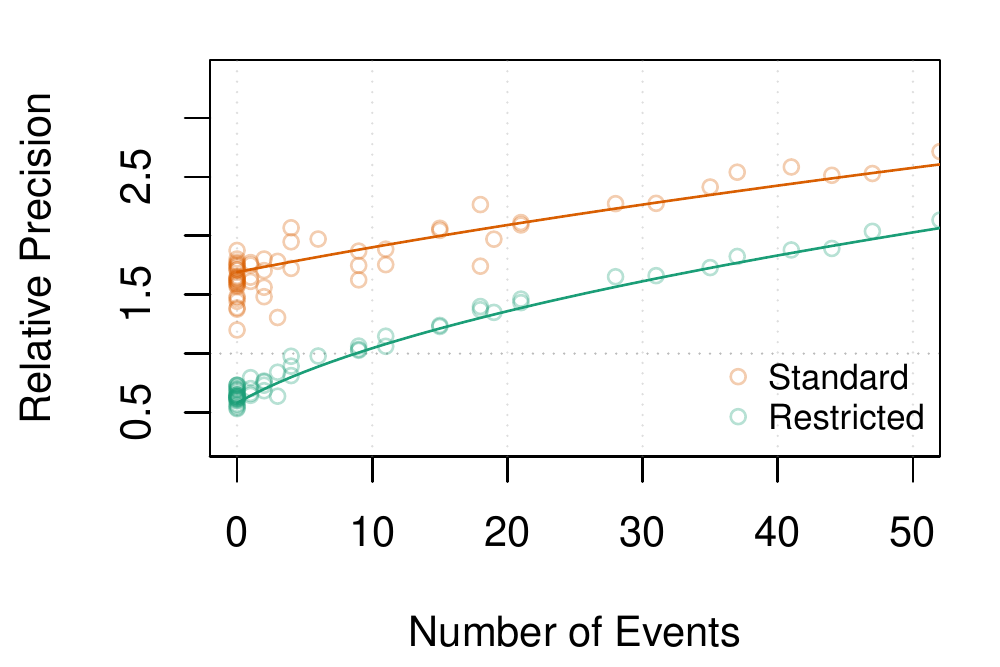}\label{fig:rprec_black}}\\
        \subfigure[Asian Mothers: 2019]{\includegraphics[width=.45\textwidth]{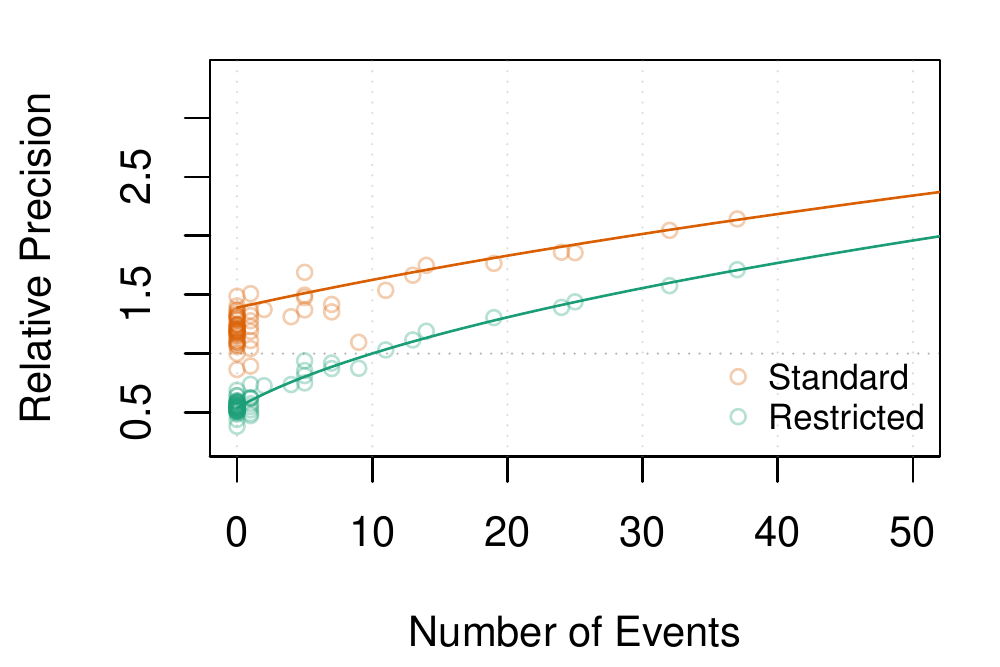}\label{fig:rprec_asian}}
        \subfigure[Hispanic Mothers: 2019]{\includegraphics[width=.45\textwidth]{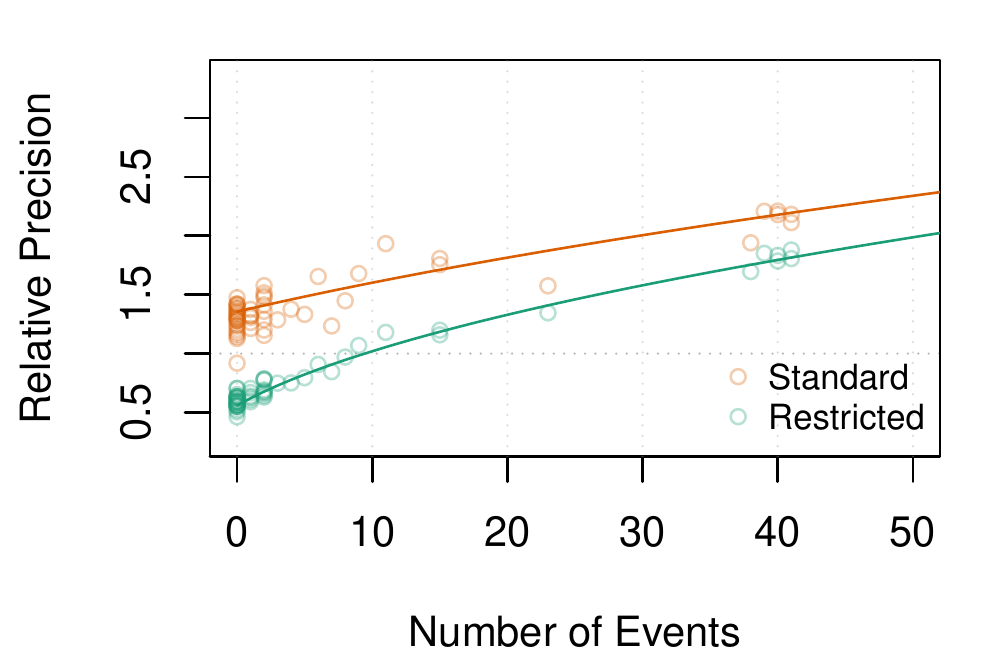}\label{fig:rprec_hisp}}
    \end{center}
    \caption{Comparison of the relative precision of the race/ethnicity-specific estimates from 2019 under the standard and restricted CAR models.}
    \label{fig:relprec}
\end{figure}

Having discussed the effect oversmoothing can have on an individual county's estimates, we now shift our attention to the primary focuses of this paper: maps of estimates and reliability.  To do so, we consider the maps shown in Figure~\ref{fig:ratemaps} and the relative precision plots shown in Figure~\ref{fig:relprec}. We begin by comparing the estimates for white mothers in 2019 based on the standard CAR model in Figure~\ref{fig:white_rate_full} to those based on the restricted CAR model in Figure~\ref{fig:white_rate_thres}.  While the overall geographic patterns are similar, the estimates from the restricted model have more extreme values, a manifestation of the phenomenon observed in Figure~\ref{fig:white_hist}.  In addition, estimates for several counties are deemed unreliable in the restricted model; in contrast, all of the estimates produced by the standard CAR model are deemed reliable.  As shown in the relative precision plot in Figure~\ref{fig:rprec_white}, the unreliable estimates under the restricted model (i.e., those with a relative precision less than 1) all correspond to instances where $y_{irt}<10$, demonstrating that our $\widehat{a}_{0;rt}<5$ restriction had the desired effect.

While the results for white mothers in Figures~\ref{fig:white_rate_full} and~\ref{fig:white_rate_thres} provided a \emph{subtle} illustration of the difference between the standard and restricted CAR models, the results for Asian mothers in Figures~\ref{fig:asian_rate_full} and~\ref{fig:asian_rate_thres} illustrate a much more stark difference.  Specifically, unlike Pennsylvania's white population, racial minorities in Pennsylvania are much more geographically concentrated in the state's major metropolitan areas.  As such, the heightened informativeness of the standard CAR model not only yields reliable estimates in all but two counties, but it also produces a very spatially smooth map as very few counties experienced enough preterm births to overrule the model's spatial structure.  In contrast, the restricted CAR model again results in a \emph{de facto} requirement that $y_{irt}\ge 10$ to obtain a reliable estimate (as shown in Figure~\ref{fig:rprec_asian}), producing unreliable estimates in 56 of the state's 67 counties.  Similar results are observed for black and Hispanic mothers and for the entire ten-year study period, as shown in Figures~C.1 and~C.2 of the Web Appendix.

\begin{figure}[t]
    \begin{center}
        \subfigure[Asian Mothers: 2019 (Standard)]{\includegraphics[width=.4\textwidth]{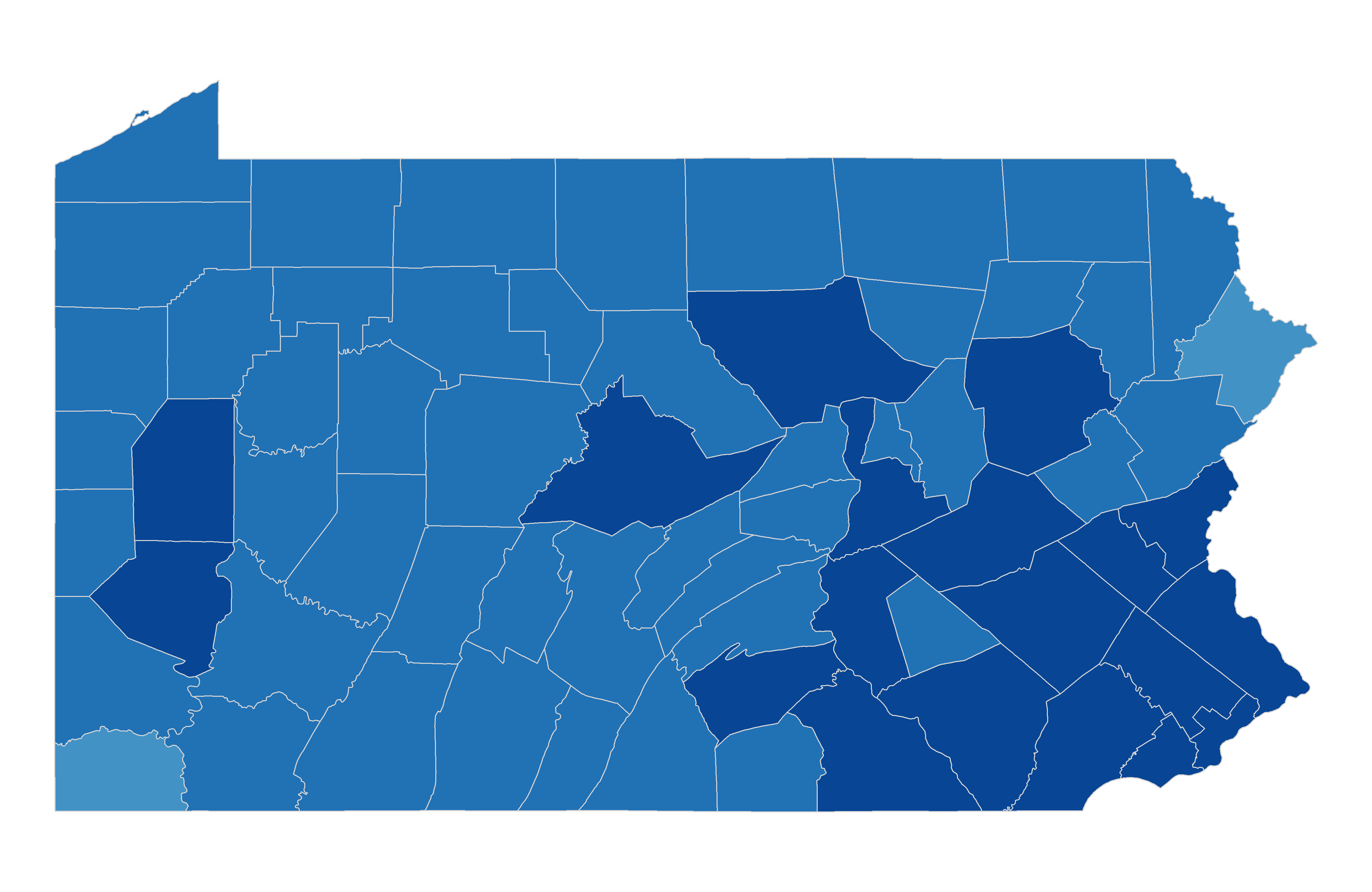}\label{fig:asian_rel_full}}
        \subfigure[Asian Mothers: 2019 (Restricted)]{\includegraphics[width=.4\textwidth]{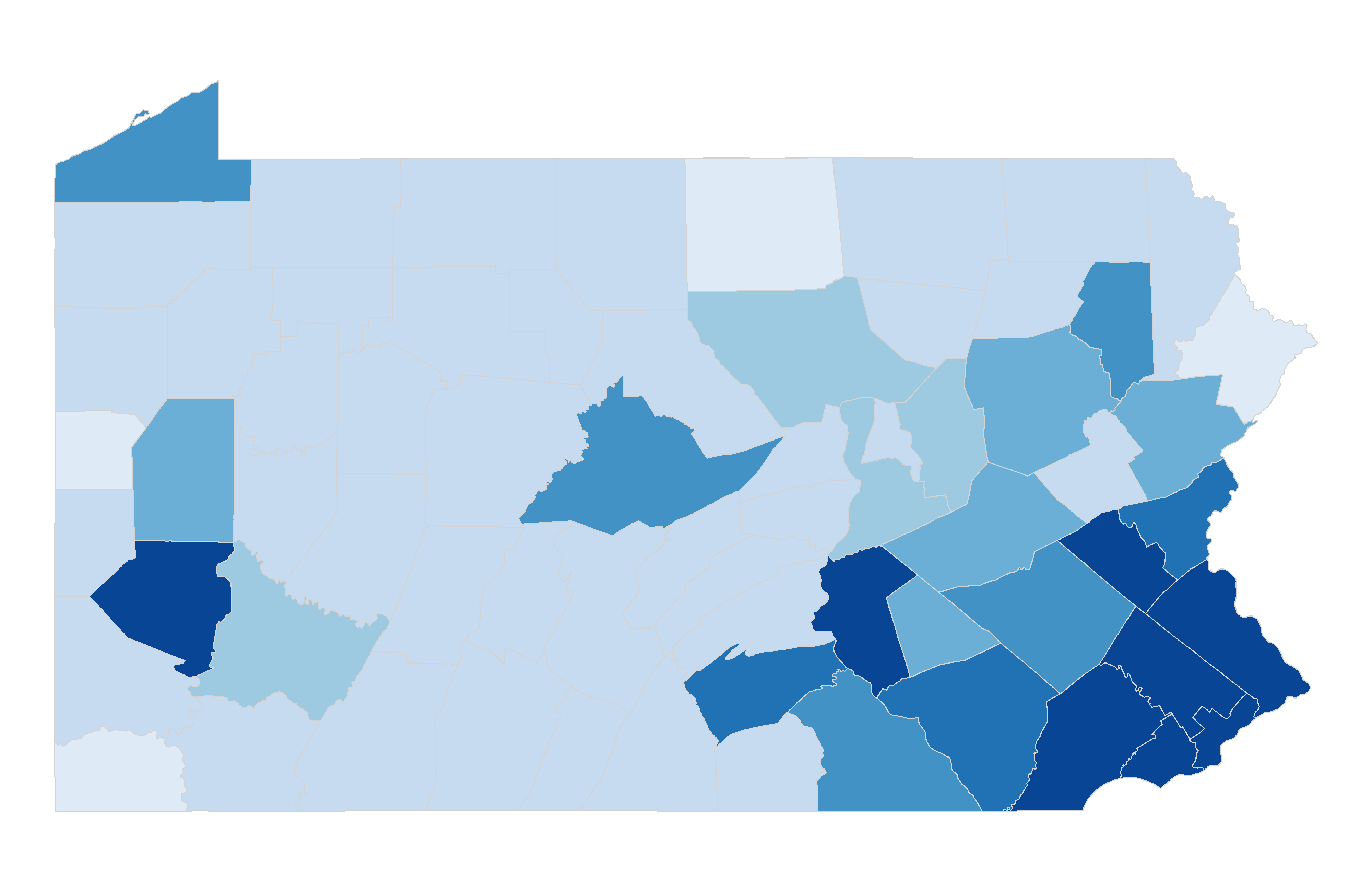}\label{fig:asian_rel_thres}}
        \includegraphics[width=.16\textwidth]{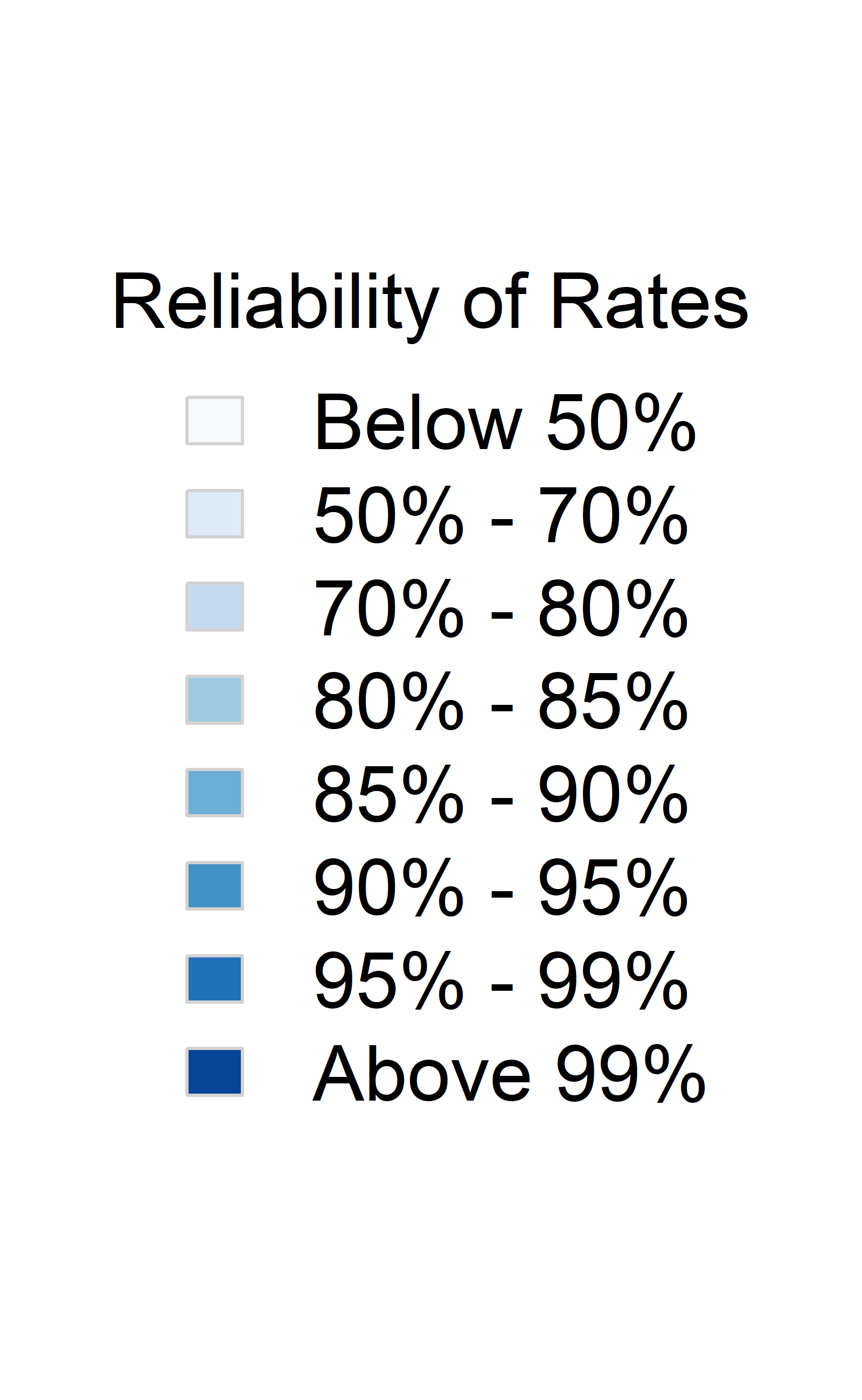}
    \end{center}
    \caption{Comparison of the level of reliability of the estimates of the preterm birth rate for Asian mothers in 2019 under the standard and restricted CAR models.}
    \label{fig:relmaps}
\end{figure}

Finally, the discussion up until this point has treated ``reliability'' as a binary ``reliable vs.\ unreliable'' property of an estimate, but a key feature of our definition of a reliable estimate is that 
we can determine an estimate's \emph{level} of reliability by identifying the value of $\alpha\in\left(0,1\right)$ such that its posterior median will be greater than the width of its $\left(1-\alpha\right) \times 100\%$ equal-tailed credible interval.  For instance, Figure~\ref{fig:relmaps} displays maps of the level of reliability of the estimates for Asian mothers in 2019 from the standard and restricted CAR models.  As first observed in Figure~\ref{fig:asian_rate_full}, Figure~\ref{fig:asian_rel_full} shows that all but two counties achieved a level of reliability greater than the 0.95 level under the standard CAR model.  In contrast, Figure~\ref{fig:asian_rel_thres} highlights the degree of geographic clustering of Pennsylvania's Asian population by virtue of the disparate levels of reliability in the estimates.  Reliability maps for other racial/ethnic groups and other years are shown in Figure~C.3 of the Web Appendix.  There, not only do we observe the stark differences in the level of reliability produced by the standard and restricted CAR models, but we also observe differences in the year-to-year variability.  In particular, while the restricted model produces reliability levels that are consistent year-to-year for each racial/ethnic group --- reflecting a similar degree of consistency in the temporal behavior of the underlying data --- the level of reliability of the estimates produced by the standard CAR model exhibits irregular behavior. This manifests in requirements of the data to achieve reliability that differ year-to-year, as can be observed in Figure~C.1 of the Web Appendix.

\section{Discussion}\label{sec:disc}
This paper was motivated by the lack of consensus in the statistical and epidemiologic literature regarding the requirements for an estimate of an event rate to be deemed ``reliable''.  The proposed definition in Section~\ref{sec:methods} accommodates both crude and model-based estimates, as well as discrete (reliable vs.\ unreliable, based on some predetermined level) and continuous statements of reliability.  Equally important, our definition of reliability can be directly related to the \emph{posterior} number of events --- i.e., the sum of the \emph{observed} number of events and the \emph{prior} number of events --- thereby allowing users to restrict the informativeness of their model specification such that a minimum number of events must be observed in order to expect to obtain a reliable estimate.  Moreover, while properties of our definition of reliability are most clearly conveyed via the conjugate beta-binomial and Poisson-gamma modeling frameworks, the approximations proposed in \citet{bym:info} and \citet{song:bin} allow us to extend these properties to the \citet{bym}-framework commonly used in the disease mapping literature where the question of reliability often arises.  Future work aims to extend these restrictions to other, more recently developed approaches for disease mapping \citep[e.g.,][]{leroux:car,datta:dagar} and methods for multivariate spatial and spatiotemporal settings \citep[e.g.,][]{gelfand:mcar,nonsep:poisson}, particularly for the purpose of estimating age-standardized rates from age-stratified event data.

A principal underpinning of this work is that declaring an estimate as ``reliable'' conveys an element of \emph{trustworthiness}, not simply that an estimate is \emph{precise}.  As such, we believe the motivation of using model-based estimates should not be to produce ``reliable'' estimates but rather to improve the estimates' precision in a deliberate and measured manner.  Thus, the objective of this paper is to describe how researchers can exercise restraint and use criteria for reliability to inform the design of their statistical models to obtain estimates that have a desired level of precision.  In addition, a key feature of the reliability definition provided here is that the continuous quantification of reliability provides researchers some flexibility with regard to presenting their results.  For instance, Figure~C.4 of the Web Appendix illustrates how dynamic and/or interactive data visualization tools can be used to display estimates with varying levels of reliability as an alternative to producing maps with few ``reliable'' estimates or to artificially inflating the precision of estimates (e.g., by using an overly informative model) for the purpose of having more estimates being deemed ``reliable''.

We would also be remiss to acknowledge that there are situations where the all but the most highly populated regions will not experience enough events in a standard amount of time (e.g., one year) to be deemed reliable, with or without the contribution of prior information, particularly when producing estimates for and making inference on rare outcomes.  In these situations, rather than increasing the model's informativeness, the requirements for reliability can instead be used as a basis for combining data across multiple time periods.  For instance, based on the expected preterm birth rates for each of the race/ethnicities considered in Section~\ref{sec:analysis}, combining four years of data with a model specification that contributed $\widehat{a}_0 \approx 5$ prior events would be sufficient to produce reliable estimates for one-third of Pennsylvania's counties for Asian mothers and for half of Pennsylvania's counties for both black and Hispanic mothers, improvements of more than 10 counties for each group compared to using a single year of data.  As with the notion of using a relaxed level of reliability, however, agencies would need to weigh the benefits of improving reliability via aggregation versus producing estimates that might be considered outdated before they have even been published.

\bibliographystyle{jasa}
\bibliography{chr_ref, lbw}

\end{document}